\documentclass[twocolumn,showpacs,amsmath,epsfig,apsrev]{revtex4}

\usepackage{graphicx}
\usepackage{graphics}
\usepackage{dcolumn}
\usepackage{bm}

\begin{document}

\title{Neutron spectroscopy and magnetic relaxation of the Mn$_6$ nanomagnets.}

\author{S. Carretta$^1$, T. Guidi$^2$, P. Santini$^1$, G. Amoretti$^1$,O. Pieper$^{3,4}$, B. Lake$^{3,4}$, J. van Slageren$^{5,6}$,
F. El Hallak$^{5}$, W. Wernsdorfer$^7$, H. Mutka$^8$, M.
Russina$^3$,  C.J. Milios$^9$ and E. K. Brechin$^9$}

\address{$^1$Dipartimento di Fisica, Universit\`{a} di Parma, I-43100 Parma, Italy}

\address{$^2$ ISIS Pulsed Neutron Facility, Rutherford Appleton Laboratory, Chilton, OX11 0QX, United Kingdom}

\address{$^3$ Hahn-Meitner Institut, Glienicker Strasse 100, 14109 Berlin, Germany}

\address{$^4$ Institut f\"{u}r Festk\"{o}rperphysik, Technische Universit\"{a}t Berlin, 10623 Berlin, Germany}

\address{$^5$ 1. Physikalisches Institut, Universit\"{a}t Stuttgart, D-70550 Stuttgart, Germany}

\address{$^6$ School of Chemistry, University of Nottingham, Nottingham NG7 2RD, United Kingdom}


\address{$^7$ Laboratoire Louis N\'{e}el-CNRS, F-38042 Grenoble Cedex, France}

\address{$^8$ Institute Laue-Langevin, B.P. 156, F-38042 Grenoble Cedex, France}

\address{$^9$ University of Edinburgh, West Mains Road, Edinburgh, EH9 3JJ, U.K.}

\begin{abstract}
Inelastic neutron scattering has been used to determine the
microscopic Hamiltonian describing two high-spin variants of the
high-anisotropy Mn$_6$ nanomagnet. The energy spectrum of both
systems is characterized by the presence of several excited
total-spin multiplets partially overlapping the S=12 ground
multiplet. This implies that the relaxation processes of these
molecules are different from those occurring in prototype
giant-spin nanomagnets. In particular, we show that both the
height of the energy barrier and resonant tunnelling processes are
greatly influenced by low-lying excited total-spin multiplets.
\end{abstract}


\date{\today}
\maketitle

 \vspace{1cm}

\section{Introduction}
Molecular nanomagnets (MNMs) have recently attracted considerable
interest because at low temperature $T$ they display slow
relaxation of the magnetization $M$ of purely molecular origin
\cite{Gatteschibook}. The main relaxation mechanism is provided by
the interactions of the spin degrees of freedom with phonons,
either through modulation of the local crystal fields on
individual magnetic ions, or through modulation of two-ion
interactions. Typically, the relaxation dynamics are modelled by
restricting the spin Hilbert space to the ground total-spin
multiplet only, as resulting from the usually dominating isotropic
exchange interactions. In this framework, each $N$-spins molecule
is described as a single giant spin $S$ in an effective
crystal-field potential. At temperatures of few K, the reversal of
$M$ occurs through a multi-step Orbach process yielding a
thermally activated behavior of the relaxation time, $\tau =
\tau_0 exp(U/k_BT)$, where the energy barrier $U$ is set by the
effective axial anisotropy experienced by the giant spin\cite{Villain}.\\
We show that excited $S$-multiplets strongly influence the energy
barrier for the relaxation of $M$ in two high-spin ($S = 12$)
variants of the high-anisotropy Mn$_6$ nanomagnet (a record
barrier $U = 86.4$ K, and $U = 53.1$ K \cite{Mn6high,Mn6low}). In
fact, inelastic neutron scattering (INS) and Frequency Domain
Magnetic Resonance Spectroscopy (FDMRS) show that the two variants
are characterized by a similar anisotropy but have significantly
different exchange interactions. The large difference in $U$ is
mainly due to the presence of relaxation paths passing through
excited $S$ multiplets partially nested within the ground one. In
addition, because of $S$-mixing in the wavefunctions, these
excited manifolds may lead to resonant inter-multiplet tunnelling
processes for fields of a few thousands of Gauss. These are
associated with additional steps in hysteresis
cycles which are absent in the giant spin model.\\

\section{Spin Hamiltonian and molecular energy levels}

The two Mn$_{6}$ molecules have chemical formula
Mn$_6$O$_2$(Et-sao)$_6$(O$_2$CPh(Me)$_2$)$_2$(EtOH)$_6$ (higher
barrier) and
Mn$_6$O$_2$(Et-sao)$_6$(O$_2$CPh)$_2$(EtOH)$_4$(H$_2$O)$_2$ (lower
barrier) and are nearly isostructural. The magnetic core comprises
six Mn$^{3+}$ ions arranged on two triangles bridged by oxygen
atoms (Fig. 1). Each Mn$^{3+}$ ion has a distorted octahedral cage
of ligands, with the Jahn$-$Teller axes all approximately
perpendicular to the planes of the triangles.
\begin{figure}[ht]
\includegraphics[width=8cm,angle=0]{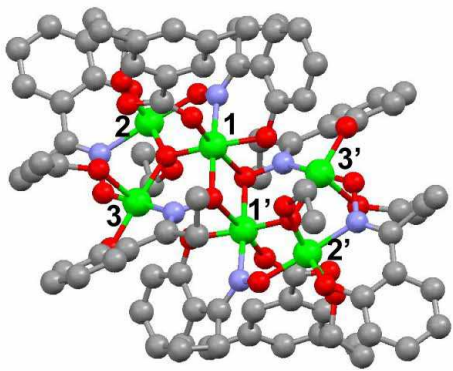}
\caption{(color online) Structure of the high-barrier Mn$_{6}$
molecule (H omitted for clarity). Green : Mn, red : O, blue : N,
dark grey : C).}
\end{figure}
The six Mn$^{3+}$ ions have spin $s$ = 2 and are coupled by
dominant ferromagnetic interactions, leading to a high $S=12$
total-spin ground state, as can be inferred by magnetization
measurements, see Fig. 2 \cite{Mn6high,Mn6low}.\\
\begin{figure}[ht]
\includegraphics[width=7cm,angle=0]{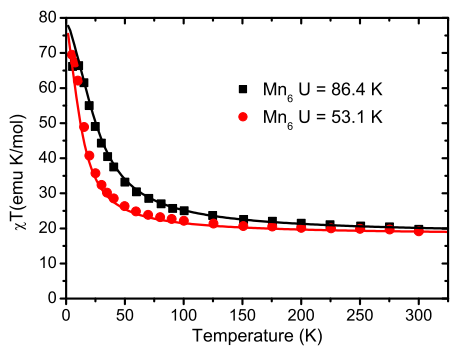}
\caption{(color online) Measured (points) and calculated (lines)
$T$-dependence of the susceptibility for the two Mn$_6$ molecules.
Calculations are made by including exchange interactions only,
with the parameters determined by INS. $g=2$ has been assumed.}
\end{figure}
Each Mn$_{6}$ molecule can be described by the following spin
Hamiltonian\cite{Mn6prl}
$$
H=\sum_{i < j}J_{i j}{\bf s}(i)\cdot {\bf s}(j) + \sum_{i} d_i
s_z^2(i) +
$$
$$
\sum_{i} c_i (35 s_z^4(i) + (25 -30 s(s+1)) s_z^2(i)) + g \mu_B
{\bf B}\cdot {\bf S} + H^\prime,\eqno(1)
$$
where ${\bf s}(i)$ are spin operators of the $i^{th}$ Mn ion and
${\bf S} = \sum_{i}{\bf s}(i)$ is the molecule's total spin. The
first term is the isotropic exchange, while the second and third
terms describe axial local crystal-fields (a $z$ axis
perpendicular to the triangles plane is assumed), and ${\bf B}$ is
the external field\cite{Gatteschibook}. $H^\prime$ (neglected in
the following) represents additional small anisotropic terms. The
minimal set of free parameters is given by three different
exchange constants $J_{11^\prime}\equiv J_1$,
$J_{12}=J_{23}=J_{13}=J_{1^\prime 2^\prime}=J_{2^\prime
3^\prime}=J_{1^\prime 3^\prime}\equiv J_2$, and
$J_{13^\prime}=J_{1^\prime 3}\equiv J_3$ (Fig.1) and two sets of
crystal-field (CF) parameters $d_1=d_{1^\prime}$,
$c_1=c_{1^\prime}$, and $d_2=d_{2^\prime}$, $c_2=c_{2^\prime}$.
The ligand cages of sites 1 and 3 are rather similar and we
assumed the corresponding CF parameters to be equal.  Since
experimental information is insufficient to fix independently the
two small $c$ parameters,we have chosen to constrain the ratio
$c_1/c_2$ to the ratio $d_1/d_2$. The anisotropic terms break
rotational invariance and here lead to a large amount of mixing of
different $S$ multiplets ($S$ mixing \cite{QTM}). In the following
we label the states by their
leading $S$-component.\\
\begin{figure}[ht]
\includegraphics[width=7.5cm]{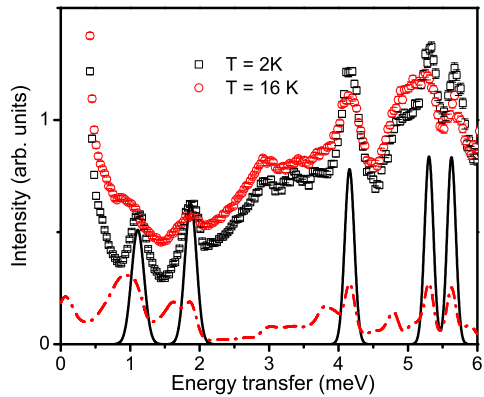}
\caption{(color online) INS spectra for the higher barrier
molecule collected on IN5 with incident wavelength of 3.4 {\AA}
for T= 2 K (black) and T = 16 K (red). Lines are theoretical
calculations with model (1).}
\end{figure}
\begin{figure}[ht]
\includegraphics[width=7.5cm]{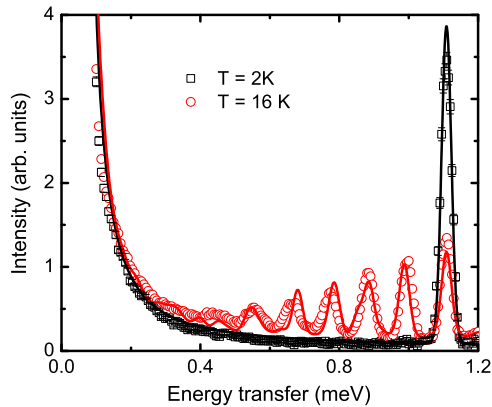}
\caption{(color online) INS spectra for the higher barrier
molecule collected on IN5 with incident wavelength of 6.7 {\AA}
for T= 2 K (black) and T = 16 K (red). Lines are theoretical
calculations with model (1).}
\end{figure}
To determine the parameters appearing in (1), we have used the
time-of-flight neutron spectrometers NEAT at the Hahn Meitner
Institut (Berlin) and IN5 at the Institute Laue Langevin
(Grenoble). In addition, we have performed FDMRS measurements at
the University of Stuttgart. Since FDMRS is sensitive to
intra-multiplet transitions only, its use in conjunction with INS
makes easier to assess the character of the different observed
excitations. Figs. 3 and 4 show examples of INS spectra together
with theoretical simulations. The analysis of INS and FDMRS data
leads to the following parameters for the higher (lower) barrier
compounds : $J_1 = -0.84 (-0.61)$ meV, $J_2 = -0.59 (-0.31)$ meV,
$J_3 = 0.01 (0.07)$ meV, $d_1 = -0.20 (-0.23)$ meV, $d_2 = -0.76
(-0.97)$ meV and $c_1 = -0.001 (-0.0008)$ meV \cite{Mn6prl}. These
parameters are consistent with susceptibility measurements (see
Fig. 2). Hence, anisotropy is similar in the two variants whereas
the dominant ferromagnetic exchange is substantially larger in the
higher-barrier compound. Figures 5 and 6 show the energies of the
$S$-multiplets resulting from the exchange part of (1). It is
evident from these figures that the exchange splitting between the
ground $S=12$ manifold and many excited multiplets, including
low-$S$ ones, is smaller than the energy scale of anisotropic
terms in (1). In particular, Fig. 6 shows that in the
lower-barrier compound the energy of the lowest-lying $S=0$
multiplet is only about 4 meV larger than that of the ground one.
\begin{figure}[ht]
\includegraphics[width=7.5cm]{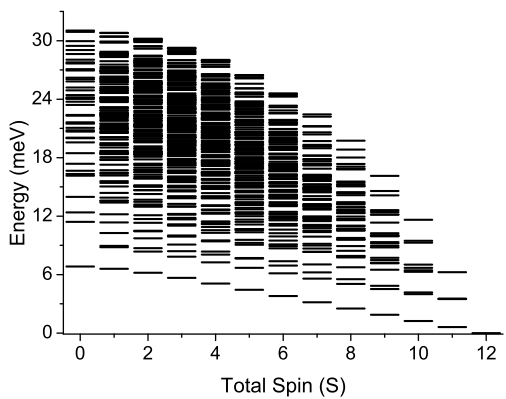}
\caption{Energy of all $S$-multiplets for the higher barrier
compound, resulting from the exchange part of (1) with the
parameters given in the text.}
\end{figure}
\begin{figure}[ht]
\includegraphics[width=7.5cm]{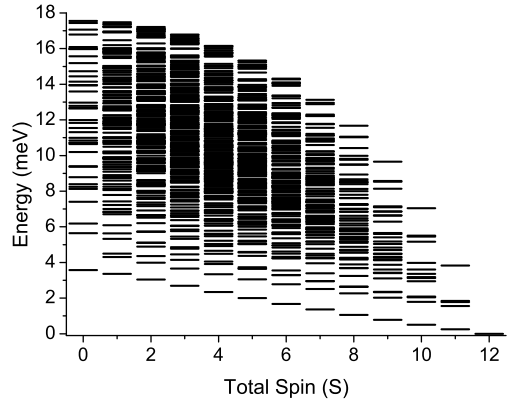}
\caption{Energy of all $S$-multiplets for the lower barrier
compound, resulting from the exchange part of (1) with the
parameters given in the text.}
\end{figure}
Thus, the giant spin mapping completely breaks down in these two
molecules, not only for the large $S$-mixing in the wavefunctions,
but even for failing to account for the number of states located
below the barrier. The presence of a center of inversion
(characterizing the structure determined at 150 K) implies that
exchange multiplets can be classified according to their parity
with respect to the associated spin-permutation operation. For
instance, the ground $S=12$ states are even, whereas the
lowest-lying $S=11$ states are odd.

\section{Relaxation of the Magnetization}

The main difference between the two Mn$_6$ variants is the
position of the excited $S$ manifolds, therefore the comparison
between the relaxation behavior of these systems offers the
opportunity to study the role played by low-lying excited
multiplets. To address this issue, we adopt the well established
framework presented in \cite{Blum,nmr} for the irreversible
evolution of the density matrix $\rho (t)$. By focusing on time
scales much longer than the typical periods of free evolution of
the system, the so-called ''secular approximation'' enables the
time evolution of the diagonal matrix elements of the density
matrix to be decoupled from that of the off-diagonal ones. In
particular, the populations of the molecular eigenstates, $p_l
(t)$, evolve through master equations:
$$
\dot{p}_l(t) = \sum_m W_{l m}p_m(t), \eqno(2)
$$
where $W_{l m}$ is the $l m$ element of the rate matrix, i.e., the
probability per unit time that a transition between levels $\vert
 m \rangle$ and $\vert l \rangle$ is induced by the interaction
with the phonon heat bath. The latter can be calculated by
perturbation theory once magnetolelastic (ME) interactions have
been modelled. Experimental information is totally insufficient to
fix all the many possible parameters appearing in the ME coupling
potential. By assuming that the main contribution to this coupling
arises from the modulation of the local rank-2 crystal fields and
that quadrupole moments of each individual Mn ion are
isotropically coupled to Debye acoustic phonons, we obtain
\cite{Ni10}
$$
W_{l m}=\gamma^2 \Delta^3 _{l m} n(\Delta_{l m}) \sum_{i,j,q_1,
q_2} \langle l \vert O_{q_1,q_2}({\bf s}_i)\vert m \rangle \times
$$
$$
{\langle m \vert O_{q_1,q_2}({\bf s}_j)\vert l \rangle}, \eqno(3)
$$
where $i$ and $j$ run over Mn ions, $O_{q_1,q_2}({\bf s}_i)$ are
the components of the cartesian quadrupole tensor operator,
$n(x)=(e^{\hbar x/k_{\rm B}T}-1)^{-1}$ and $\Delta_{l
m}=(E_l-E_m)/\hbar$. The single free parameter $\gamma$ is
proportional to the ME coupling strength. In spite of the
complexity of their energy spectrum, for both variants the
resulting relaxation spectrum at low $T$ is dominated by a {\it
single} relaxation time displaying a nearly Arrhenius behavior
$\tau (T)= \tau_0 \exp(U/K_B T)$, in agreement with experiments
\cite{Mn6prl}. In particular, the effective energy barrier for the
magnetization reversal is crucially dependent on the position of
low-lying excited $S$ multiplets. Indeed, the large difference
between the barriers of the two molecules mainly results from the
variation of the ferromagnetic exchange constants and not from a
variation in anisotropy. It is also worth to note that in the
present case $U$ is not set by the energy of the lowest-lying
$M_S=0$ state as one could naively expect.\\
\begin{figure}[ht]
\includegraphics[width=8.5cm]{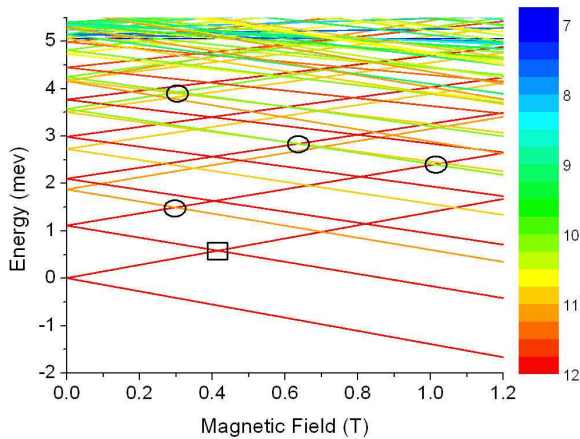}
\caption{Energy levels of the higher-barrier compound as a
function of $B$, applied along the $z$ axis. The color maps for
each state the value of $S_{eff}$, where
$<S^2>:=S_{eff}(S_{eff}+1)$. Ellipses and the square indicate
representative examples of level crossings.}
\end{figure}
The existence of several total-spin multiplets partially
overlapping with the ground one implies that a magnetic field
applied along the molecule's easy axis induces crossings involving
states which are absent in a giant spin description. These can be
turned into anticrossings (ACs) by the small transverse terms
contained in $H^\prime$. Figure 7 shows that besides the
''traditional" intra-$S=12$ crossings at about 0, 0.4 (e.g., the
square) and 0.8 T, many more occur for intermediate values of $B$.
In particular, there are three different kinds of
''non-traditional" field-induced crossings, i.e., which are absent
in the giant spin description of the molecule. First of all there
are crossings involving a pair of states belonging to a single
low-energy $S$ multiplet different from the ground one (for
instance the higher-lying ellipsis at 0.3 T in Fig. 7). These
crossing may lead to relaxation shortcuts if the dominant
relaxation path passes through them. The other two kinds of
crossings (e.g. the other ellipses in Fig. 7) involve pairs of
states belonging to different $S$-manifolds. The distinction
arises from the symmetry properties of the two crossing states,
i.e., they may have the same parity or not. In the first case
(e.g., the crossings between red and green curves in the two
ellipses at about 0.6 and 1 T) $H^\prime$ usually leads to an AC,
whereas in the second case (e.g., the ellipse at about 0.3 T) an
AC may occur to the extent that at low $T$ the inversion center is
removed by a distortion, leading to terms with low-enough symmetry
in $H^\prime$. Even if there are no structural data below 150 K,
it is not unlikely for magnetic molecules to undergo a symmetry
lowering at low $T$ (see, e.g., \cite{cr8,Fe4}).
''Non-traditional" ACs with the associated resonant incoherent
tunneling, may result in relaxation shortcuts leading to
additional steps in hysteresis cycles, absent in a giant-spin
picture. For instance, Fig. 8 shows an example of derivative of
the hysteresis curves measured at $T = 3$ K for the higher-barrier
molecule\cite{Mn6high}. The ''traditional" ACS produce minor
features in these curves, and the two main peaks are associated
with ''non-traditional" ACs. The effect of low-field ACs is more
evident at higher $T$\cite{Mn6prl}. For instance, there are
features at 0.3 T which may a priori originate from both crossings
marked on Fig. 7.
\begin{figure}[ht]
\includegraphics[width=6.5cm]{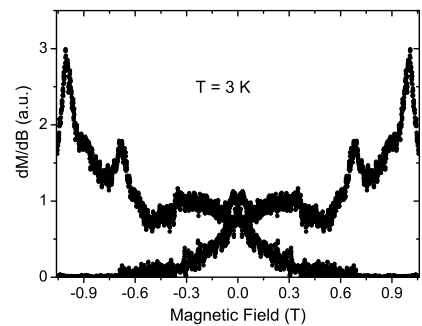}
\caption{Example of derivative of the hysteresis curves measured
for the higher-barrier molecule in \cite{Mn6high}. For each value
of $B$, there are two points corresponding to increasing or
decreasing $B$ in the hysteresis cycle.}
\end{figure}

\section{Conclusions}

By exploiting INS and FDMRS we have determined the microscopic
Hamiltonian of two different high-spin variants of the Mn$_6$
nanomagnet. We have found that excited $S$ multiplets overlapping
with the ground one strongly affect the magnetic relaxation
process.  Moreover, we have demonstrated the existence of
tunnelling pathways involving pairs of states belonging to
different total spin manifolds. Hence, the energy separation
between the ground and excited multiplets may be a key ingredient
in determining the relaxation and tunnelling dynamics of molecular
nanomagnets.

\end{document}